\documentstyle[aps]{revtex}
\begin{document}
\voffset=-15mm

\title{Double fluctuations on the attractive
Hubbard\\ model: ladder approximation}
\author{S. Schafroth}
\address{Physik-Institut der Universit\"at Z\"urich, 
Winterthurerstrasse 190, CH-8057 Z\"urich, Switzerland.}
\author{J.J. Rodr\'{\i}guez-N\'u\~nez}
\address{Instituto de F\'{\i}sica, Universidade Federal Fluminense,\\ 
Av.\ Litor\^anea S/N, Boa Viagem, \\
24210-340 Niter\'oi RJ, 
Brazil. \\e-m: jjrn@if.uff.br}
\date{\today}
\maketitle

\begin{abstract}

We explore, for the first time,  
the effect of double fluctuations on both the diagonal and 
off-diagonal self-energy.  
We use the T-Matrix equations below $T_c$, developed recently by 
the Z\"urich group  
(M.H. Pedersen et al)   
for the local pair attraction Hamiltonian. 
Here, 
we 
include as well the effect of fluctuations on the order parameter (beyond 
the BCS solution) up to second order 
in $U/t$. This is equivalent to approximating 
the effective interaction by 
$U$ in the off-diagonal self-energy. 
For $U/t = -6.0$, $T/t = 0.05$, $\mu/t = - 5.5$ and $\Delta/t = 1.5$, 
we find four peaks both for the diagonal, $A(n(\pi/16,\pi/16),\omega)$, and 
off-diagonal, $B(n(\pi/16,\pi/16),\omega)$, spectral functions. These peaks are 
not symmetric in pairs as previously found. In addition: ({\bf a}) 
in $A(n(\pi/16,\pi/16),\omega)$, 
the far left peak has a vanishing small weight; 
({\bf b}) in $B(n(\pi/16,\pi/16),\omega)$ 
the far left and far right peaks have 
very small weights. The physical picture 
is, then, that the pair physics in the normal phase ($T > T_c$) is still 
valid below $T_c$. However, the condensation of the e-h pairs produces an 
additional gap around the chemical potential as in $BCS$, 
in other words, superconductivity opens a gap in the lower 
branch of a Hubbard-type-I solution.
\\
Pacs numbers: 74.20.-Fg, 74.10.-z, 74.60.-w, 74.72.-h
\end{abstract}



\section{Introduction}

High-temperature superconductors display a wide range of behaviour
atypical of the standard band-theory of metals. In the superconducting state
these materials become extreme type II superconductors, with a 
short coherence volume, which one might take as an indication of
bound pairs. These features indicate that correlation effects
might be important in understanding the physical nature of these
materials. One of the simplest models featuring superconductivity and
allowing a systematic study of the effect of electron correlations is
the attractive Hubbard model. Although this model is unlikely to be a
microscopic model of high-temperature superconductors, it is likely
that understanding it will provide  insights into the effect of
correlations on measurable properties. We adopt the local pairing 
potential in 2-D as a starting point to study the effect of correlations 
(beyond a simple $BCS$ approach) due to the fact that the $HTSC$ materials 
do not seem to be described by a mean field approach, as it was 
recognized by Randeria et al\cite{Randeria}. Furthermore, Puchkov, 
Basov and Timusk\cite{PBT} pointed out that in the $HTSC$, the Fermi 
surface is estimated to be $E_F = 1-2~ev$, which is not much larger 
than the energies probed in infrared experiments ($4-300~meV$). Such a 
low $E_F$ could be a reason for violation of a quasi-particle 
description, i.e., we do not have well-defined elementary excitations. 
This fact may require the use of the local potential with correlations, and 
indeed, the damping obtained from numerical calculations is comparable 
to the Fermi energy, for small electron concentration.

In a previous work the effect of electron correlations on some normal-state 
properties of the attractive Hubbard model was studied using a selfconsistent 
T-matrix formalism, going beyond simple mean-field treatments\cite{tmatrix}.  
It was found that
for intermediate coupling strengths 
the attractive interaction gives rise to
large momentum  bound states with energies below the
two-particle continuum and with a pronounced effect on the spectral 
properties. Namely a splitting 
of the free band into two, one of which is associated
with virtual bound states.  Furthermore, a bending in the static
spin-susceptibility was observed for temperatures just above the
phase-transtion\cite{TMJJ}.

In a following paper\cite{singleflu}, the group of R\"uschlikon 
used the functional derivative formalism\cite{BK}. They derived  
the T-Matrix equations in the superconducting phase valid up to 
second order in the off-diagonal one-particle Green functions. The 
main feature of that communication was that the T-Matrix,  
appeared both in the 
diagonal (first order in the $T$-Matrix) and off-diagonal (order parameter 
plus a second order contribution in 
the $T$-Matrix) part of the self-energy.   
The authors of Ref.\cite{singleflu} implemented a low order 
approximation with full fluctuations in the diagonal part of the 
self-energy while using a mean field approximation for the 
off-diagonal part of the self-energy, i.e., equal to the order 
parameter.

In this communication, we implement a higher order approximation to 
the off-diagonal self-energy by allowing fluctuations in the order 
parameter also, i.e., making $T({\bf x,x}') \approx U \delta({\bf x-x'})$. 
We keep the full $T$-Matrix in the diagonal part of the self-energy as 
in Ref.\cite{singleflu}. The goal of the 
present work is to study the stability of the 
physical picture given in Ref.\cite{singleflu}, which 
can be summarized as the appareance of four peaks, symmetric in 
pairs, both in $A(k,k,\omega)$ and $B(k,k,\omega)$. We 
caution the reader that we are exploring the effect of 
double fluctuations by approximating $T({\bf x,x}')$ 
by $U\delta({\bf x-x'})$ in $\Sigma_{12}({\bf x,x}')$ as a step 
forward in taking into account the full effect of 
correlations. In Section \ref{eqsmot} 
we present the model and the results of Ref.\cite{singleflu}. 

In Section \ref{Lowest-order} we keep the 
low order approximation taken in Ref.\cite{singleflu} and 
implement our 
second order approximation in the off-diagonal self-energy. 
Our implementation is performed with the fast-Fourier transform 
(FFT) and we disscuss the physical meaning of our 
results. Section \ref{conclus} concludes.

\section{The model and the T-Matrix equations.}\label{eqsmot}

The Hubbard Hamiltonian is defined as
\begin{eqnarray}
\label{Ham}
    H = - t\sum_{<ll'>\sigma}c_{l\sigma}^{\dagger}c_{l'\sigma}
   + U \sum_l n_{l\uparrow}n_{l\downarrow}
   - \mu \sum_{l\sigma} n_{l \sigma}~~,
\end{eqnarray}
where the $c_{l\sigma}^{\dagger}$($c_{l\sigma}$) are creation (annihilation)
operators for electrons 
with spin $\sigma$.  The number-operator is $n_{l\sigma}
\equiv c_{l\sigma}^{\dagger}c_{l\sigma}$, $t$ is the hopping matrix element
between nearest neighbours $l$ and $l'$, $U$ is the onsite interaction
and $\mu$ is the chemical potential in the grand canonical ensemble.
Here we consider an attractive interaction, $U<0$. For a review
of the attractive Hubbard model see Micnas et al\cite{Micnas_et_al}.   
Previous authors have 
used this model to study the bismuthate  superconductors\cite{5}.

By starting with the Nambu Green function 
\begin{equation}\label{Green}
{\cal G}_{{\bf x x}'}(\tau , \tau') \equiv
        - \left< T_{\tau}\left[ \Psi_x(\tau) \otimes
        \Psi_{x'}^{\dagger}(\tau') \right] \right> ~~~,
\end{equation}
\noindent
where $\otimes$ means the tensor product and $T_{\tau}$
means the time ordering of the two Nambu operators, where these are 
defined, at position ${\bf x}$ and imaginary time, $\tau$, by
\begin{equation}\label{spinors}
 \Psi_{\bf x} = \left( \begin{array}{ll}
                     c_{{\bf x} \uparrow} \\
                        c_{{\bf x} \downarrow}^{\dagger}
                   \end{array}    \right)  ~~~;
~~~\Psi_{\bf x}^{\dagger} = \left(
                     c_{{\bf x} \uparrow}^{\dagger} ~~~
                         c_{{\bf x} \downarrow}
                    \right)~~~~
\end{equation}
the authors of Ref.\cite{singleflu} write the solution to Dyson's 
equation in $\bf k,\omega$-space as

%
%
%

\begin{eqnarray} \label{solu}
    G_{11}({\bf k},i\omega_n)  = - G_{22}({\bf k},-i\omega_n) =  
	~~~~~~~~~~~~~~~~~~~~~ \nonumber \\
        \frac{i\omega_n + \varepsilon_{\bf k} -
        \Sigma_{22}({\bf k},i\omega_n)}{(i\omega_n - \varepsilon_{\bf k} -
        \Sigma_{11}({\bf k},i\omega_n))(i\omega_n + \varepsilon_{\bf k} -
        \Sigma_{22}({\bf k},i\omega_n)) - \Sigma_{12}({\bf k},i\omega_n)
        \Sigma_{21}({\bf k},i\omega_n)} ~ , 
\end{eqnarray}
\begin{eqnarray} \label{solu2}
    G_{12}({\bf k},i\omega_n) = G_{21}({\bf k},-i\omega_n) = 
	~~~~~~~~~~~~~~~~~~~~~ \nonumber \\
        \frac{\Sigma_{12}({\bf k},i\omega_n)}{(i\omega_n - 
	\varepsilon_{\bf k} -
        \Sigma_{11}({\bf k},i\omega_n))(i\omega_n + \varepsilon_{\bf k} -
        \Sigma_{22}({\bf k},i\omega_n)) - \Sigma_{12}({\bf k},i\omega_n)
        \Sigma_{21}({\bf k},i\omega_n)} ~ ,
\end{eqnarray}
\noindent
where $\omega_n \equiv \pi (2n+1)/\beta$ are the
fermionic Matsubara frequencies and $\beta=1/(k_BT)$ the inverse
temperature. The $d$-dimensional dispersion is given by
$\varepsilon_{\bf k} = -2t \sum_{\alpha=1}^d \cos(k_\alpha a_\alpha)-\mu$. 
$\Sigma$ is the self-energy matrix\cite{Mattuck} and $d$ is the lattice 
dimension.  

Then, by using the functional derivative techique\cite{BK}, the authors 
of Ref.\cite{singleflu} get the self-energies 
to second order in $G_{12}(G_{21})$
\begin{eqnarray} \label{self-energies}
    \Sigma_{11}({\bf x,x}') &=& G_{22}({\bf x,x}')T({\bf x',x}) + 
        G_{21}({\bf x},\overline{a}) 
	G_{22}(\overline{a},\overline{b})T(\overline{b},{\bf x})
            G_{12}(\overline{b},{\bf x'})T({\bf x',x}) ~~~, \nonumber \\
    \Sigma_{12}({\bf x,x}') &=& \Delta({\bf x}) \delta({\bf x-x'}) + 
	G_{22}({\bf x},\overline{a})G_{12}(\overline{a},\overline{b})
        T(\overline{b},{\bf x})G_{11}(\overline{b},{\bf x'})
        T({\bf x'},\overline{a})~~~, \nonumber \\
    \Sigma_{21}({\bf x,x}') &=& 
	\Delta'({\bf x}) \delta({\bf x -x'}) + 
	G_{11}({\bf x},\overline{a})G_{21}(\overline{a},\overline{b})
        T(\overline{b},{\bf x})G_{22}(\overline{b},{\bf x'})
        T({\bf x'},\overline{a})~~~, \nonumber \\
    \Sigma_{22}({\bf x,x}') &=& G_{11}({\bf x,x}')T({\bf x',x}) + 
         G_{12}({\bf x},\overline{a})G_{11}
        (\overline{a},\overline{b})
	T(\overline{b},{\bf x})G_{21}(\overline{b},{\bf x'})T({\bf x',x}) ~~~.
\end{eqnarray}
where $T({\bf x,y})$ in reciprocal space has the following form
\begin{equation} \label{T-matrix}
    T(q,i\varepsilon_m) = \frac{U}{1-U \chi(q,i\varepsilon_m)}~~~~,
\end{equation}
with
\begin{equation} \label{suscept}
     \chi(q,i\varepsilon_m) \equiv \frac{1}{N\beta}
    \sum_{k,i\omega_n} 
	G_{22}(k-q,i\omega_n-i\varepsilon_m)G_{11}(k,i\omega_n)~~~,
\end{equation}
\noindent
and $\varepsilon_n \equiv 2 
\pi n/\beta$ are the bosonic Matsubara frequencies. 
This solution (see Eq. (\ref{self-energies})) 
is valid for $\Delta/W \ll 1$, where $W=2dt$ is the bandwidth. 
In Eq. (\ref{self-energies}) summation over repeated indices 
is understood (space and imaginary time). The physical meaning of 
Eq. (\ref{self-energies}) is the repeated scattering of two 
particles many times, without polarization of the medium. This is  
equivalent to the $T-$Matrix approximation which is valid for 
small carrier concentrations. Another approximation involved in 
the derivation of Eq. (\ref{self-energies}) is the condition 
that $\Delta(T)~<<~W$. This is equivalent to a perturbation 
expansion to second order in $\Delta(T)/W$. 
For a solution of the Eqs.\ 
(\ref{solu},\ref{T-matrix},\ref{suscept},\ref{self-energies})
one would also need to fix the 
chemical potential from the particle number using
\begin{equation} \label{number}
    \rho(T,\mu) = \lim_{\eta \rightarrow 0^+}
 \frac{1}{\beta N}\sum_{\omega_n,{\bf k}}G({\bf k},i\omega_n)
        exp(i\omega_n \eta),
\end{equation}
\noindent where $\rho$ is the electron concentration 
per spin and is defined in the interval $[0,1]$.  
Thus, the set of Eqs.\
(\ref{solu},\ref{T-matrix},\ref{suscept}
,\ref{number})
represents a set of non-linear 
self-consistent equations which needs to be solved
numerically.  

We note that an expansion of the final equations, 
Eqs.~(\ref{self-energies}), to first order in $U$ simply
gives the wellknown BCS expressions. To second in $U$ the result is
identical to that of Mart\'{\i}n-Rodero and Flores\cite{M-R&F}. 

\section{Numerical results} \label{Lowest-order}

Before we disclose our approximation and present a numerical 
solution to it, we will discuss the approximation made in 
Ref.\cite{singleflu}. The authors of this reference approximate 
the set of Eqs. (\ref{self-energies}) in the following form
\cite{Rodriguez,Beckapp} 

\begin{eqnarray}\label{singlefluc}
\Sigma_{11}({\bf x,x}') &=& G_{22}({\bf x,x}')T({\bf x',x}) \nonumber \\
\Sigma_{12}({\bf x,x}') &=& \Delta({\bf x})\delta({\bf x-x'}) \nonumber \\
\Sigma_{21}({\bf x,x}') &=& \Delta^*({\bf x})\delta({\bf x-x'}) \nonumber \\ 
\Sigma_{22}({\bf x,x}') &=& G_{11}({\bf x,x}')T({\bf x',x}) ~~,  
\end{eqnarray}

\noindent
which is equivalent to keeping full fluctuations in the diagonal 
part of the self-energy while performing a mean field approximation 
in the off-diagonal self-energy, 
$\Sigma_{12}({\bf x,x}') = \Delta({\bf x})$ 
$\delta({\bf x-x'})$. Remember  
fluctuations 
enter through the $T$-Matrix, and this has been set equal to 
zero in the off-diagonal self-energy. The idea behind the point of 
view taken by these authors is that the effect of correlations is 
mainly present in the diagonal self-energy, i.e., correlations are 
linear in the $T$-Matrix. In addition, the authors of Ref.\cite{singleflu} 
took the pragmatic view of taking one step at a time, i.e., to study 
correlations in the same manner as they were studied in the normal 
state.

Here, we want to include fluctuations in the order parameter, or 
equivalently in the off-diagonal part of the self-energy. For this, 
we make $T({\bf x,x}') \approx U \delta({\bf x - x'})$ in the off-diagonal 
self-energy while keeping the full fluctuations in the diagonal 
self-energy. Then, our approximation is equivalent to the 
following set of equations

\begin{eqnarray}\label{doublefluc}
\Sigma_{11}({\bf x,x}') &=& G_{22}({\bf x,x}')T({\bf x',x}) \nonumber \\
\Sigma_{12}({\bf x,x}') = \Delta({\bf x})\delta({\bf x-x'}) &+& 
U^2 G_{11}({\bf x,x}') G_{12} ({\bf x',x}) 
G_{22}({\bf x,x}')\nonumber \\
\Sigma_{21}({\bf x,x}') = \Delta^*({\bf x})\delta({\bf x-x'}) &+& 
U^2 G_{22}({\bf x,x}') G_{21} 
({\bf x',x}) G_{11}({\bf x,x}') \nonumber \\ 
\Sigma_{22}({\bf x,x}') &=& G_{11}({\bf x,x}')T({\bf x',x})  
\end{eqnarray}
which are local in real space where they can be easily evaluated 
numerically. The Greens functions are determined 
selfconsistently using Dyson's
equation where the order parameter is used as 
input (see below). The technical aspects of the numerical
solution using the FFT-technique has been detailed in \cite{tmatrix}. 
We consider that the set of Eqs. (\ref{doublefluc}) contains double 
fluctuations since we have full fluctuations in the diagonal 
part of the self-energy and fluctuations in the off-diagonal 
self-energy. The latter come through the one-particle 
Green functions themselves which must be calculated 
self-consistently. The reason that we pursue the present approximation 
is that   
we believe that including fluctuations in the off-diagonal self-energy 
will have a strong influence on the dynamical properties. The 
outcome of our numerical calculations vindicate this. Although the 
choice of $T({\bf x,x}') \approx U \delta({\bf x - x'})$ in the 
off-diagonal self-energy does not fully take into account all the 
fluctuations, it keeps the numerical implementation under 
control with present time Workstation facilities. We leave to the 
future an implementation of the full $T$-Matrix contribution. 

The numerical simulations were performed in two dimensions (2D), for 
$U/t=-6.0$, $\mu = - 5.5$, $\Delta = 1.5$ and $T/t=0.05$.  
We have fixed the order parameter and the chemical potential 
because computationally is much easier to find one root rather than two. 
The selected value of $\Delta/t = 1.5$ for $U/t = -6.0$ is inside 
the range of validity of the approximation and close to the 
numerical value found in Ref.\cite{singleflu}. As is explained in 
Ref.\cite{S-RN-B}  numerical convergence is
based on the fact that the diagonal self-energy  
depends neither on $\mu$ nor $\Delta$. This assumption is not longer 
valid here, making the program more complex. Because of these 
reasons, we have decided to fix $\Delta$ and $\mu$ and find 
the carrier concentration, $\rho$. 
In addition, there is 
more than one solution for the equation $\rho(\Delta,\mu) = 
const$ (see Fig. 8).  
Here one must note that due to the Mermin-Wagner theorem no
phase-transition is expected in 2D systems with a continous
symmetry, 
and the formalism employed is therefore too
simple to describe a Kosterlitz-Thouless phase-transition.
Nonetheless, it is observed that the formalism does give a
phase-transition, in agreement with the results above the transition
temperature, where the signaling of 
a divergence of the T-matrix was observed.

	In Figure 1, we present the diagonal one-particle 
spectral function, $A({\bf k},\omega)$, defined by
\begin{equation}
    A({\bf k},\omega) \equiv
     - \frac{1}{\pi} \lim_{\delta \rightarrow O^{+}}
     Im[G_{11}({\bf k},\omega + i\delta)],
\end{equation}
The value of the gap, $\Delta/t = 1.5$ is within the region
of validity of the expansion, 
Eqs.~(\ref{self-energies}), i.e., 
$\Delta/W < 1$, with $W = 8t$ the 
bandwidth in 2D. We observe three visible peaks, two symmetric 
around the chemical potential, $\mu$, and the third is the the 
{\it upper} Hubbard branch. However, there is a fourth peak for 
$\omega < 0$, which for $n = 0$ is around $\omega \approx - 3.8$. The 
equivalent peak for $\omega > 0$ is around $\approx + 3.25$.  
The peak at $\omega \approx - 3.8$ has a vanishingly small weight (see the 
inset figure). So, we conclude that the symmetric peak structure does 
not hold anymore (at least for the most extreme peaks). If we 
consider that the peak at $\omega \approx - 3.8$ is indeed small 
then we are left with effectively three peaks.  
These results are different from the ones 
obtained by the Z\"urich group\cite{singleflu}. Then, the picture 
which emerges from $A({\bf k},\omega)$ is the appearence of 
three energy branches, a higher one which is already present in the 
normal phase and two lower ones which are symmetric around the 
chemical potential. These two superconducting lower branches 
correspond to the lower energy branch of the normal state. 
Now superconductivity ($\Delta \neq 0$) opens up a gap in the 
normal lower energy branch. These two superconducting low branches 
are, then, due to the pairing of the electron - hole pairs accross the 
chemical potential, similar to the $BCS$ case. The only difference
with respect to the weak coupling ($BCS$) limit is the appearence 
of the higher energy branch. 

	In Figure 2, we present the off-diagonal one-particle 
spectral function, $B({\bf k},\omega)$, which is defined as,   
\begin{equation}
    B({\bf k},\omega) \equiv  - \frac{1}{\pi} 
    \lim_{\delta \rightarrow O^{+}}
    Im[G_{12}({\bf k},\omega + i\delta)],
\end{equation}
From Figure 2 we see that this function 
has two symmetric and visible peaks. In the inset 
we show a blow up of the energy 
scale. We observe two additional
two peaks at $\omega 
\approx + 3.9, - 4.1$, for $n = 0$,  which are not 
symmetric and their weights are small. So that we argue that they 
can be neglected. Finally, there are two symmetric peaks around 
$\mu$ ($\omega = 0$). The 
conclusion to be drawn from these observations is that the off-diagonal 
spectral function remains qualitatively equal to the the $BCS$ 
case, i.e., the inclusion of fluctuations both in the diagonal 
and off-diagonal components of the Nambu self-energy does not 
change the basic $BCS$ condensation picture. Only the diagonal 
spectral function suffers the effect of correlations (Figure 1), 
as should be, due to the presence of strong correlations. 
We have chosen $k_x$ and $k_y$ in units of 
$\pi/16$.

	In Figure 3 we present the imaginary part 
of the two-particle Green's
function, $-ImG_2({\bf q},\omega)$. 
The Cooper resonance almost dissapears since the peak weights for 
$\omega < 0$ are negligible (see inset). In addition, for 
every value of $k$ we find ($\omega > 0$) a main peak with some two 
broad peaks of reasonable weights. These additional broad peaks 
have an effect on $\Sigma({\bf k},i\omega_n)$. 
As $\Sigma({\bf k},i\omega_n)$,  
$G({\bf k},i\omega_n)$ and $F({\bf k},i\omega_n)$ are 
coupled together, then, these peaks which in the 
are going to have an effect on the quasi-particle spectra. All 
this is due to the fact that our equations are self-consistent. 
We observe a well defined two-particle structure for all values 
of momentum for $\omega > 0$. The fact that the 
Cooper resonance is, in certain sense, washed out is 
probably due to the fact 
that the effect of double fluctuations favor pairing formation,  
which implies that the interaction is 
stronger. A discussion of this point is treated in the paper 
of Ranninger and Robin\cite{RR}. Another difference 
between our results and those of  
the Z\"urich group, is that our band of two particles is wider 
than theirs. This has deep 
physical consequences, as we will see later. 

	In Figure 4 we present the imaginary part of the diagonal 
self-energy, $-Im[\Sigma(n(\pi/16,\pi/16),\omega)]$ 
for the same parameters 
of Figure 3.  
We observe an opening of the gap at the chemical 
potential, $\omega = 0$. This gap is much bigger than twice 
the order parameter. In the inset we have a blow up of the 
energy scale concluding that it is composed of two peaks, one 
of them broad. The authors of Ref.\cite{tmatrix,TMJJ} found that 
$Im\Sigma$ almost has a peak in the normal 
phase. The presence of this peak is the origin  
to the two peaks in $A(n(\pi/16,\pi/16),\omega)]$. In 
the superconducting phase, the authors of Ref.\cite{singleflu} 
find that $Im\Sigma$ still has one peak which implies the 
presence of four peaks, symmetric in pairs, both in 
$A(n(\pi/16,\pi/16),\omega)]$ and $B(n(\pi/16,\pi/16),\omega)]$.  
Now, in our approach with double fluctuations, we observe that 
$Im\Sigma$ has no well defined single peak structure. This, we
argue, will wash out the pair symmetry. The presence of 
{\it double fluctuations} is responsible for the cancellation of 
the leftest peak of $A(n(\pi/16,\pi/16),\omega)]$. So, the picture 
which emerges is a Hubbard-type-I solution combined with a 
superconducting gap in the lower branch\cite{H1}.

	In Figure 5 we show the real part of 
the diagonal self-energy, $Re[\Sigma(n(\pi/16,\pi/16),\omega)]$, 
for the same parameters 
given in Figure 3. We observe in Figure 4 that the width of these 
peaks is bigger than those of the Z\"urich group. The consequence 
of this is that we are no longer able to define a well define 
peak in the self-energy, so that the four peak structure (with 
symmetric pair of peaks) of the 
Z\"urich group is not longer valid. Thus, the picture is the 
following: above $T_c$ we have two peaks which are due to the 
pair formation physics, i.e., the {\it upper} and {\it lower}  
branches; below $T_c$ one of the bands is split into two 
to the spontaneous symmetry breaking, 
$\Delta \neq 0$. 

	In Figures 6 and 7 we show the frequency dependence of 
the real and imaginary part of the off-diagonal self-energy, 
i.e., $Re[\Sigma_{12}(n(\pi/16,\pi/16),\omega)]$ 
and $-Im[\Sigma_{12}(n(\pi/16,\pi/16),\omega)]$, 
for the same parameters as given previously. 
For the case of $-Im[\Sigma_{12}(n(\pi/16,\pi/16),\omega)]$ we have 
not shown the full energy range. However, it is 
fully antisymmetric. To the best of our 
knowledge, this is the first presentation of these results in 
the literature. We observe that $Im[\Sigma_{12}(n(\pi/16,\pi/16),\omega)]$ and 
$Im[\Sigma(n(\pi/16,\pi/16),\omega)]$ have almost the same values for large 
frequencies. However, the position of their peaks is at a different 
frequency for each $n$. In addition, 
the effect of correlations (fluctuations) 
is important since they decrease the value of 
$Re[\Sigma_{12}(n(\pi/16,\pi/16),\omega)]$ from $\Delta$ 
down to almost $\Delta/2$, 
for $n = 0$ and $\omega = 0$. If we take a closer look at 
Eqs.(\ref{solu},\ref{solu2}) we see that $\Sigma_{12}$ plays the role of 
an energy gap. Then, the energy gap is (${\bf k},\omega$)-dependent 
and it of the same order of magnitude than the order parameter, 
$\Delta(T)$. This analogy leads us to conclude that the order parameter, 
$\Delta(T)$, and the energy gap are two different quantities. So, a 
local interaction produces a (${\bf k},\omega$)-dependent energy 
gap, which is the quantity that experimentalists are most likely 
measuring.

	In Figure 8 we have a plot of $\rho$ vs $\mu$ for 
different fixed values of the order parameter, $\Delta$. We 
see that there is an abrupt change of density as a function 
of $\mu$. Also, for a fixed value of $\rho$ there are up to 
three different values of the chemical potential. Then, our scheme 
is not valid for large densities, a condition we take as 
granted by the definition of the $T$-Matrix formalism. 
On the other hand, when we increase the temperature the step 
of $\rho$ vs $\mu$ tends to dissapear. This implies that our
 scheme is valid for temperatures which are not 
too low. We would like 
to point out that the implementation of the $T$-Matrix approach 
in the normal phase $(\Delta = 0$) did not converge for small 
temperatures and high densities\cite{tmatrix}. 
We interpret the non-convergence 
of the $T-$Matrix program above $T_c$ as due 
the fact that we are approaching 
a parameter region where there are no physical solutions. For 
example, in another context, Figueira and Foglio\cite{FF}, have 
found multivalues for the chemical potential in the periodic 
Anderson model and they 
choose the solution of minimun 
Helmholtz free energy.  

\section{Conclusions.}\label{conclus}

Using the analytical 
results of Ref.\cite{singleflu}, we have implemented 
a program where we have included 
fluctuations both in the diagonal and off-diagonal 
self-energy. This approach we call double fluctuation since 
we have full fluctuations in the diagonal part of the self-energy 
while going beyond the mean field solution ($BCS$) in the order 
parameter. This represents a higher order 
approximation than those used in Ref.\cite{singleflu}. We have 
presented the one-particle and two-particle spectral functions 
pointing out the differences with the low order approximation 
performed by the Z\"urich group\cite{singleflu}. We have indicated that the 
pair formation, which is a phenomenon valid even above $T_c$ remains 
valid in our scheme. However, the effect of fluctuations in the 
off-diagonal part of the diagonal self-energy ({\it double 
fluctuations}) does not produce a well defined peak in the 
diagonal part of the one-particle self-energy and this breaks 
the four peak structure found by the Z\"urich group, resulting 
in a clearer picture since only one of the Hubbard bands is split into 
two around the chemical potential. The off-diagonal spectral 
density is qualitatively similar to the one of $BCS$, but the 
diagonal spectral function suffers the effect of strong correlations. 
As has been said before, the physical picture which emerges from 
our work is that of superconductivity opening a gap in the lower 
branch of a Hubbard-type-I solution. 
We have performed additional calculations and found that the opening 
of the correlation gap is hard to achieve.  
For $U/t = - 12.0$, $T/t = 0.1$ and 
$\mu/t = - 8.5$ we find that the correlation gap is more or 
less visible but not fully developed. This is due to the fact that 
the shifting of the electronic states around the chemical potential 
sends electronic states above $\mu$ and this closes the correlation gap. 
We have also discussed the parameter space where our approximation 
is valid and we have made a connection with the calculations above 
the transition temperature discussing the validity of 
each.  
To end, we mention that a previous calculation by Rodr\'{\i}guez - 
N\'u\~nez, Cordeiro and Delfino\cite{14} has used the ideas presented 
in this work within the framework of 
the moment approach of Nolting\cite{15} in 
the superconducting phase.

\section{Acknowlegments.}
We would like to thank Brazilian Agency CNPq (Project 
No. 300705/95-96), CONDES-LUZ and CONICIT 
(Project No. F-139). Very useful discussions with Prof. M.S. 
Figueira and Prof. Gerardo Mart\'{\i}nez are fully acknowledged. 
We thank Dr. A. Umerski and 
Prof. Mar\'{\i}a Dolores Garc\'{\i}a - Gonz\'alez for reading 
the manuscript. 




\vspace{.5cm}
\newpage
\begin{center}
{\large Figures.}
\end{center}
\begin{figure}
\caption{
The diagonal one-particle spectral function, $A(n(pi,pi),\omega)$ vs
$\omega$ for different momenta along the diagonal of the 
Brillouin zone $({\bf
k} = (n,n)\pi/16)$ for $U/t = -6.0$, 
$T/t = 0.05$, $\Delta = 1.5$ and $\mu = - 5.5$.  
We have used an external
damping of $\delta/t = 0.1$.  We have 
used $16 \times 16$ points in the 
Brillouin zone and 1024 Matsubara 
frequencies. After self-consistent 
calculation of the coupled
non-linear equations, we obtain  a density, $\rho = 0.035942$. we have 
runned our source code in single precision requiring $18~MB$ of RAM 
memory. Each iteration takes 2.5 minutes of CPU time.
}
\end{figure}

\begin{figure}
\caption{
The off-diagonal one-particle spectral function, 
$B(n(\pi/16,\pi/16),\omega)$ vs
$\omega$ for different momenta along 
the diagonal of the Brillouin zone.  Same
parameters as in Figure\ 1.}
\end{figure}

\begin{figure}
\caption{
$-Im[G_2(m(\pi/16,\pi/16),\omega)]$ vs $\omega$ for different momenta along the
diagonal of the Brillouin zone (${\bf q} = (m,m)\pi/16$).  
Same parameters as in
Figure\ 1.}
\end{figure}

\begin{figure}
\caption{
$-Im[\Sigma(n(\pi/16,\pi/16),\omega)]$ vs 
$\omega$ for different momenta along the 
diagonal of the Brillouin zone (${\bf k} = (n,n)\pi/16$). 
Same parameters as in
Figure\ 1.} 
\end{figure}

\begin{figure}
\caption{
$Re[\Sigma(n(\pi/16,\pi/16),\omega)]$ 
vs $\omega$ for different momenta along the
diagonal of the Brillouin zone (${\bf k} = (n,n)\pi/16$).
Same parameters as in
Figure\ 1.}
\end{figure}

\begin{figure}
\caption{
$-Im[\Sigma_{12}(n(\pi/16,\pi/16),\omega)]$ vs $\omega$ 
for different momenta along the
diagonal of the Brillouin zone (${\bf k} = (n,n)\pi/16$).
Same parameters as in
Figure\ 1.}
\end{figure}

\begin{figure}
\caption{
$Re[\Sigma_{12}(n(\pi/16,\pi/16),\omega)]$ vs $\omega$ 
for different momenta along the
diagonal of the Brillouin zone (${\bf k} = (n,n)\pi/16$).
Same parameters as in
Figure\ 1.}
\end{figure}

\begin{figure}
\caption{
$\rho$ vs $\mu$ for different fixed values of the order 
parameter, $\Delta$, at $T/t = 0.05$ and $T/t = 0.2$.}
\end{figure}

\begin{references}

\bibitem{Randeria}
	Mohit Randeria, Nandini Travedi, Adriana Moreo and 
	Richard T. Scalettar, Phys. Rev. Lett. {\bf 69}, 
	2001 (1992).
\bibitem{PBT}
	A.V. Puchkov, D.N. Basov and T. Timusk, Cond-Matter/9611083. In 
	this paper, the authors study the presence of the pseudogap in 
	$HTSC$ by means of infrared measurements, which in our view 
	points toward the need of strong correlations; see, also, 
	A.V. Puchkov, P. Fournier, D.N. Basov, T. Timusk, A. Kapitulnik 
	and N.N. Kolesnikov, Phys. Rev. Lett. {\bf 77}, 3212 (1996). 
	In this last paper, the authors study the frequency-dependent 
	scattering rate $1/\tau(\omega,T)$ and its doping dependence. 
	While they observe a normal-state gaplike depression in 
	$1/\tau(\omega,T)$ in the underdope regime, it is not 
	observed in the overdoped regime. This points to the fact 
	that the high-$T_c$ cuprates are strongly correlated materials 
	in the underdoped regime.
\bibitem{tmatrix}
        R.\ Micnas, M.H.\ Pedersen, S.\ Schafroth, T.\ Schneider, 
    	J.J.\ Rodr\'{\i}guez-N\'u\~nez and H.\ Beck, 
    	{\em Phys.\ Rev.\ B}, {\bf 52}, 11223 (1995). 
\bibitem{TMJJ}
	T. Schneider, M.H. Pedersen and J.J. Rodr\'{\i}guez-N\'u\~nez 
	Z. Phys. B {\bf 100}, 263 (1996); J.M. Singer, M.H. Pederson, 
	T. Schneider, H. Beck and H.-G. Matuttis, Phys. Rev. B 
	{\bf 54}, 1286 (1996). 
\bibitem{singleflu}
	M.H. Pedersen, J.J. Rodr\'{\i}guez-N\'u\~nez, H. Beck, 
	T. Schneider and  
	S. Schafroth. (Z. Physik B, 1996). Accepted.
\bibitem{BK}
    	Kadanoff and Baym, {\em Quantum Statistical Mechanics}. (Advanced
   	Book Classics, Addison-Wesley Publishing Company 1989);
    	G.\ Baym and Leo P.\ Kadanoff, {\em Phys.\ Rev.\ } 
	{\bf 124}, 287 (1961);
    	G.\ Baym, {\em Phys.\ Rev.\ }  {\bf 127}, 1391 (1962).
\bibitem{Micnas_et_al}
        R.\ Micnas, J.\ Ranninger and S.\ Robaszkiewicz,
        {\em Rev.\ Mod.\ Phys.\ } {\bf 62}, 113 (1990).
\bibitem{5}
 	M. Rice and L. Sneddon, Phys. Rev. Lett. {\bf 47}, 689 (1981); 
	A.S. Alexandrov, J. Ranninger and S. Robaszkiewicz, Phys. 
	Rev. {\bf 33}, 4526 (1986).
\bibitem{Mattuck}
        Richard D.\ Mattuck, {\it A Guide to Feynman Diagrams
        in the Many-body Problem}. Dover (1992). Eqs.\ (10.18) and
        (15.58).
\bibitem{Fresard}
        R.\ Fr\'esard, B.\ Glaser and P.\ W\"olfle, 
        {\em J.\ Phys.: Condens.\  Matter} {\bf 4}, 8565 (1992).
\bibitem{M-R&F}
	A.\ Mart\'{\i}n-Rodero and F.\ Flores, 
        {\em Phys.\ Rev.\ B } {\bf  45},  13008 (1992).
\bibitem{Rodriguez}
        J.J.\ Rodr\'{\i}guez-N\'u\~nez,
        S.\ Schafroth, H.\ Beck, T.\ Schneider, 
        M.H.\ Pedersen and R.\ Micnas,
        {\em Physica B} {\bf 206-207}, 654 (1995).
\bibitem{Beckapp}
    	J.J.\ Rodr\'{\i}guez-N\'u\~nez, S.\ Schafroth, R.\ Micnas,
    	T.\ Schneider, H.\ Beck and M.H.\ Pedersen, 
    	{\em J.\ Low Temp.\ Phys.\ } {\bf 99}, 315 (1995).
\bibitem{S-RN-B} S. Schafroth, J.J.\ Rodr\'{\i}guez-N\'u\~nez and 
	H. Beck, (J. Phys.: Condens. Matter, 1996). Submitted.
\bibitem{RR}
	J. Ranninger and J.M. Robin, {\em Sol. State Commun.} {\bf 98}, 
	559 (1996).
\bibitem{H1}
	J. Hubbard, Proc. Roy. Soc. A {\bf 276}, 238 (1963).
\bibitem{FF}
	M.S. Figueira and M.E. Foglio, {\em Physica A} {\bf 208}, 279 (1994).
\bibitem{14}
	J.J. Rodr\'{\i}guez-N\'u\~nez, C.E. Cordeiro and 
	A. Delfino, {\em Physica A} (1996, at press).
\bibitem{15}
      	W. Nolting,  
      	{\em Z. Physik} {\bf 225}, 25 (1972); 
        W. Nolting, {\it Grundkurs: Theoretische
        Physik. 7 Viel-Teilchen-Theorie.} Verlag Zimmermann-Neufang
        (Ulmen -1992).
\end{references}
\end{document}